\documentclass[useAMS,usenatbib]{mn2e}

\usepackage{amsmath, amssymb}
\usepackage[dvips]{graphicx}
\usepackage{epsfig}

\newcommand{\aap}{A\&A}
\newcommand{\mnras}{MNRAS}

\newcommand{\apj}{ApJ}
\newcommand{\apjl}{ApJL}
\newcommand{\apjs}{ApJS}

\newcommand{\nat}{Nature}

\def\psrb{PSR~B1259-63}

\title[Striped pulsar winds in gamma-ray binaries] {Implication of the
striped pulsar wind model for gamma-ray binaries} 

\author[J.  P\'etri, G. Dubus]{J\'er\^ome P\'etri$^1$ and Guillaume
Dubus$^2$ \\
$^1$Observatoire Astronomique de Strasbourg, UMR 7550 Universit\'e de Strasbourg, CNRS, 11 rue de l'Universit\'e, 67000 Strasbourg, France\\
$^2$UJF-Grenoble 1 / CNRS-INSU, Institut de Plan\'etologie et d'Astrophysique de Grenoble (IPAG) UMR 5274, Grenoble, F-38041, France }

\date{Accepted . Received ; in original form \today}

\pagerange{\pageref{firstpage}--\pageref{lastpage}} 

\pubyear{2010}

\begin{document}

\label{firstpage}

\maketitle

\begin{abstract}
  Gamma-ray binaries are massive stars with compact object companions
  that are observed to emit most of their energy in the gamma-ray
  range.  One of these binaries is known to contain a radio pulsar,
  PSR~B1259-63. Synchrotron and inverse Compton emission from
  particles accelerated beyond the light cylinder in striped pulsar
  winds has been proposed to explain the X-ray to high energy (HE, $>$
  100 MeV) gamma-ray emission from isolated pulsars.  This pulsar
  model extends naturally to binary environments, where seed photons
  for inverse Compton scattering are provided by the companion star.
  Here, we investigate the possibility of gamma-ray emission from
  PSR~B1259-63 in the framework of the striped pulsar wind model.  The
  orbital geometry of PSR~B1259-63 is well constrained by observations
  and the double radio pulse suggests an almost orthogonal rotator so
  that the solid angle covered by the striped region is close to
  $4\pi$.  We calculate the orbital and rotational phase-resolved
  spectral variability and light-curves to expect.  We find that the
  recent detection by the {\em Fermi}/LAT of PSR B1259-63 close to
  periastron can be explained by a striped wind if the emission arises
  from a large range of radii ($\geq 1000 r_L$). We constrain the
  particle density number at the light-cylinder $n_L \approx 7 \times
  10^{15} \textrm{ m}^{-3}$.  The re-brightening a month after
  periastron passage could be due to interaction with additional seed
  photons from the trailing pulsar wind nebula. Striped winds may also
  be at work in the gamma-ray binaries LS I +61$\degr$303 and LS 5039,
  both of which have HE gamma-ray spectra reminiscent of those of
  pulsars and fluxes modulated on the orbital period. Gamma-ray
  pulsations are expected.  Some gamma-ray binaries should be faint in
  HE gamma rays (HESS J0632+057) because the line-of-sight does not
  cross the striped wind region.
\end{abstract}


\section{Introduction}

Pulsars are thought to be highly magnetized rotating neutron stars.
Many of these pulsars, previously known as radio emitter, have now
also firmly been detected in gamma rays \citep{2010ApJS..187..460A}.
For some of them, the radiated energy is mainly released in the
MeV/GeV range: there are called gamma-ray pulsars. Unprecedented
access to this energy range is currently provided by the {\em Fermi
Gamma-ray Space Telescope} Large Area Telescope.  With more than
fifty pulsars identified so far, and regular new identifications,
compared to the handful previously known, the {\em Fermi}/LAT has made
it possible to characterize the high energy phase-resolved spectra
(HE, $\ga$ 100 MeV) of a significant population of pulsars. These HE
spectra are fitted by a hard power law, with photon index $\Gamma$
between 1 and 2, supplemented by an exponential cut-off at energies
claimed to lie between 1 and 5~GeV \citep{2010ApJS..187..460A}. Pulsed
HE emission is usually explained by some radiation processes within
the magnetosphere of the pulsar. However, the absence of a
super-exponential cutoff and strong gamma-ray opacities argue against
models in which the gamma-ray emission arises near the surface of the
neutron star surface (`polar cap') \citep{2009ASSL..357..521H},
favoring models in which the HE emission arises from curvature
radiation of particles accelerated in the vacuum gaps of the outer
part of the magnetosphere (`outer gap') \citep{2009ASSL..357..481C}.

Pulsed HE emission may also arise in the pulsar wind well beyond the
surface of the light-cylinder \citep{2002A&A...388L..29K}. 3D
numerical modeling of pulsar magnetospheres in the force-free
approximation show that pulse profiles are better reproduced if the
emission is assumed to arise in current sheets just outside of the
light cylinder, along the separatrix \citep{2010ApJ...715.1282B}. The
field structure beyond the light-cylinder, where the rotation of the
neutron star requires that magnetic field lines open up, tends towards
a split monopole configuration. The misalignment of the magnetic
moment with respect to the rotation axis imprints a spiral wave shape
to the underlying current sheet
\citep{1990ApJ...349..538C,1994ApJ...431..397M}. The resulting {\em
striped pulsar wind} expands radially at relativistic speeds with a
period equal to that of the pulsar.  Beamed emission from
relativistically hot particles flowing in the current sheets, mostly
in the first few stripes, leads to pulsed gamma-ray emission.  In
particular, this model has been successfully applied to reproduce
observations of the phase-resolved optical polarization of the Crab
pulsar \citep{2005ApJ...627L..37P}, phase-resolved spectra of Geminga
and Vela \citep{2008sf2a.conf..259P,2009A&A...503...13P} as well as
light-curves of various {\em Fermi} pulsars
\citep{2011MNRAS.412.1870P} .

Unlike in the outer gap models, the radiative process producing the HE
gamma rays in the striped wind is not curvature radiation, which is
negligible outside the light-cylinder, but inverse Compton (IC)
scattering of ambient fields, typically the thermal X-ray
emission from the neutron star itself, the cosmic microwave background
or synchrotron self-Compton. Another possible source of seed photons,
for pulsars in binaries, is light from the companion (let it be a
second pulsar, a white dwarf, a low-mass or a massive star). The
photon field orientation and density change as a function of orbital
phase, affecting IC scattering in the stripes.  This can lead to
substantial variations in the pulsed gamma-ray spectra along the
orbit, something which is not expected in standard magnetospheric
models where the gamma-ray emission arises from curvature radiation
and is thus a priori impervious to the external photon field.

In this paper the expected variation of pulsed gamma-ray emission with
orbital phase is investigated using the specific case of PSR~B1259-63,
a 48~ms radio pulsar in a 3.4~year eccentric orbit around a
10~$M_\odot$ Be star (SS~2883). The system appears well-suited to test
the model since the Be star has a high luminosity, $L_\star\approx
10^{31}$~W, while the orbital separation $d$ varies by more than a
factor 10 between periastron ($d\approx 0.7$~a.u.) and apastron
($d\approx9.5$~a.u.).  Although pulsed emission from PSR~B1259-63 has
been detected in radio, only non-pulsed counterparts have been
observed in X-rays and in very high energy gamma-rays (VHE, $\ga$ 100
GeV) around periastron passage.  This emission is attributed to
radiation from particles accelerated at the termination shock of the
pulsar wind. Striped pulsar wind emission would be expected to add
another variable spectral component, peaking in HE gamma rays.
Recently, HE emission has been detected using the {\em Fermi}/LAT
\citep{2011arXiv1103.4108A,2011arXiv1103.3129T}. A first detection
coincided with the passage of the source around periastron
(mid-December 2011), where IC scattering from target photons of the
companion star should be most efficient. The flux above 100 MeV is
$\approx 10^{-7}\, \textrm{ph cm}^{-2} \textrm{s}^{-1}$ with a
power-law spectrum of photon index $\approx 2$. A second detection
occurred after a lull of about a month. The flux was about 10 times
higher than in mid-December and the spectrum was softer (photon index
$\approx 3.0$). No gamma-ray pulsation were detected.

The aim of this study is to investigate whether the HE gamma-ray
detection of PSR~B1259-63 can result from the orbital phase-dependence
of pulsed HE emission from the striped pulsar wind. The striped wind
model and related assumptions are presented in~\S\ref{sec:Model}.  The
specific application to PSR~B1259-63 is described
in~\S\ref{sec:Results}.  The relevance of our model to interpret the
recent detections of PSR~B1259-63 by {\em Fermi}/LAT and for other
gamma-ray binary systems like LS~5039 and LS~I~+61$\degr$303 is
discussed in~\S\ref{sec:Discussion} before concluding.

\section{Striped pulsar wind model}
\label{sec:Model}

The assumptions of the striped pulsar wind model, as already discussed
in several previous works \citep{2002A&A...388L..29K,
  2005ApJ...627L..37P, 2009A&A...503...13P, 2011MNRAS.412.1870P} are
briefly recapped in \S\ref{assumptions} and some specificities linked
to the binary environment are explained in \S\ref{bin}.

\subsection{General assumptions about the wind\label{assumptions}}

The rotation of the pulsar launches to first approximation a purely
radially expanding wind, isotropic in space and carrying a
$e^\pm$-pair density number falling off with distance to the center of
the pulsar~$r$ faster than $r^{-2}$ due to adiabatic cooling.  These
pairs are distributed according to a power law distribution function
of index~$p$ with minimum and maximum Lorentz factor respectively
$\gamma_{\rm min}$ and $\gamma_{\rm max}$ such that
\begin{equation}
\label{eq:FD}
dn = n(\gamma) \, d\gamma = K_e \, \gamma^{-p} \, d\gamma
\end{equation}
with~$K_e$ given by Eq.~(8) in \citet{2009A&A...503...13P} with
\begin{equation}
K_e = \frac{(N-N_0){\rm sech}^2(\Delta_\varphi \psi)+N_0}{r^2}.
\end{equation}
This expression does not yet include adiabatic cooling because it
falls off like $r^{-2}$. Nevertheless, adiabatic cooling can be
accounted for by altering the radial power law dependence of the
square of the distance~$r$ in the expression for the particle
density number. Adding an extra radial dependence, assuming a non
accelerating wind, we find a modified power law such that the
density number decreases as~$r^{-2(p+2)/3}$, the power law exponent
depends on the exact motion of the wind, linear acceleration would
lead to another exponent \citep{2002A&A...388L..29K}.  Here, the
maximal energy $\gamma_{\rm max} \, m_e \, c^2$ of the pairs (in
the comoving frame) sets the location of the cutoff in the IC
gamma-ray spectrum, an important observational constraint.  We argue
in \S3.3.3 that the characteristic $\gamma$ of this cutoff (or break)
can be related to other parameters of the striped wind.  $N_0$ sets
the background particle density number outside the current sheet
whereas $N$ represents the maximal density within the sheet.  For more
details, we refer to \cite{2009A&A...503...13P}.

The magnetic field geometry assumes a split monopole for which exact
analytical solutions exist, see for instance
\cite{1999A&A...349.1017B}. The current sheet separating the two
regions of opposite magnetic polarity remains small compared to the
wavelength of the striped wind, with a finite thickness parameterized
by the quantity $\Delta_\varphi$ (defined such that its value is high
for a thin sheet).  A smooth polarity reversal of the toroidal
component of the magnetic field $B_\varphi$ is enforced in this
transition layer (Eq.~(5)-(6) in \citealt{2009A&A...503...13P}). Note
that in the far asymptotic wind zone, the poloidal component of the
magnetic field is negligible. We enforce it strictly to zero.
Therefore, this asymptotic structure is a simple Archimedean spiral
with $B_r = B_\theta = 0$ and $B_\varphi\propto r^{-1}$.  The
structure propagates radially outwards with a constant Lorentz factor
$\Gamma_{\rm v}$ and the line-of-sight is inclined by an angle $\zeta$
with respect to the neutron star rotation axis. The sheet wiggles
around the rotational equatorial plane between angles $+\chi$ and
$-\chi$, where $\chi$ is the obliquity of the pulsar, i.e. the angle
between magnetic moment and rotation axis.  Outside the current sheet,
the plasma is cold but strongly magnetized whereas inside the sheet,
it is hot and almost unmagnetized due to the zero magnetic field point
(polarity reversal). The cold magnetized part has negligible thermal
pressure.

The pairs in the current sheet emit synchrotron (X-rays) and IC
(gamma-ray) radiation. Radiative cooling of the particles is not taken
into account but the adiabatic one is. However, only the first stripes
contribute significantly since the number density of particles
decreases faster than $r^{-2}$. The emission is boosted by
relativistic aberration when the wind expansion is directed towards
the observer.  Strong pulsed emission therefore requires that the
line-of-sight cross the current sheet i.e. that
$|\pi/2-\zeta|\leq\chi$. In this special geometric configuration the
current sheet crosses the line-of-sight twice per rotation of the
neutron star, producing two pulses of boosted emission (or only one
pulse in the less favorable case where $|\pi/2-\zeta|\approx \chi$).
The peak-to-peak phase separation of the pulses is given by
$\arccos\left(\cot \chi \, \cot\zeta\right)$, which is exactly
180$\degr$ for a line of sight contained in the equatorial plane,
independent of the obliquity. The width of the pulses depends on the
bulk Lorentz factor of the wind $\Gamma_{\rm v}$ and on the current
sheet thickness $\Delta_\varphi$. The amplitude of the pulses is set
in part by the ratio of particle density inside and outside the
current sheet. For a detailed discussion on the relation between
geometry and pulse shape, see \cite{2011MNRAS.412.1870P}.

The emission extends over a range of radii starting beyond the
light-cylinder $r_L\equiv c\, P / 2\, \pi$ for a pulsar of period~$P$.
The pulses are computed by integrating the IC emissivity over the
first few stripes, taking into account Doppler boosting, time
retardation effects and adiabatic expansion.

% \begin{figure}
% \centerline{\includegraphics[width=45mm]{sw_density_30}\includegraphics[width=45mm]{sw_density_90}}
% \caption{Log of the density of particles around the pulsar, averaged
%   over one rotation period, assuming the magnetic field obliquity is
%   $\chi=30\degr$ (left) or 90$\degr$ (right). The rotation axis of
%   the neutron star is along the y-axis; $r_L$ is the light cylinder
%   radius. The Lorentz factor of the wind is $\Gamma_{\rm v}=10$; the
%   density contrast between the current sheet and surrounding plasma
%   is $n/n_0=100$ and the thinness of the current sheet is
%   $1/\Delta_\varphi=6\degr$.}
% \label{fig:density}
% \end{figure}

The level of IC gamma-ray emission will also depend on the density of
particles in the pulsar wind. 
% This density is shown in Figure~\ref{fig:density}, averaged over the
% pulsar period. Due to symmetry by reflection about the equatorial
% plane, we only show the upper half space region where $z>0$.
The density is stronger in the striped wind region (angles between
$\pi/2$ and $\pi/2-\chi$ with respect to the vertical axis) because of
the contrast $N/N_0$ with the cold, strongly magnetized part of the
wind.  The maximum density for a given distance to the neutron star is
at an angle~$\chi$, along the crest of the current sheet wave. The
width of this maximum is tied to the thinness of the current sheet
$\Delta_\varphi$.  All other things being equal, the strongest IC
emission will occur when the line-of-sight passes through the edge of
the striped wind, in the region spanned around a latitude~$\chi$. A
single pulse in HE gamma rays is expected for this particular pulsar
orientation. A small amount of (unpulsed) IC emission may also be
expected when the line-of-sight does not cross the striped wind
because of the assumed small residual density of particles $N_0/r^2$
away from the current sheet modulo adiabatic cooling. The emission
will be unpulsed and the cold pairs will produce a sharp Compton line
\citep{2000APh....12..335B, 2007MNRAS.380..320K, 2008A&A...488...37C}
or broader band gamma-ray emission if the pairs are assumed to have
power law distributions
\citep{2005MNRAS.356..711S,2007ApJ...671L.145S,2008APh....30..239S}.
Here, the energy dissipation is limited to the striped part of the
wind, entailing the geometrical constraints described above for HE
gamma-ray observability.

\subsection{Binary environment\label{bin}}

In a binary, the orientation of the pulsar seen by the observer does
not change along the orbit (except possibly on super-orbital
timescales due to precession of the neutron star). Hence, synchrotron
emission from the striped wind stays constant. However, IC emission
from up-scattered stellar photons will change with orbital phase
because the angle between the neutron star, the companion and the
observer changes \citep{2008A&A...477..691D}. The calculation of the
gamma-ray emission is changed from previous works to correctly follow
the anisotropic scattering and to include relativistic aberration in
the pulsar wind frame of this external source of seed photons. Useful
analytical expressions can be derived for a power-law distribution of
$e^+e^-$ scattering off photons with a blackbody spectrum from a point
source, both in the Thomson \citep{2010A&A...516A..18D} and
Klein-Nishina regimes. Here we use approximate expressions for a mono
energetic photon field as given by \cite{2000ApJ...528..357M} (see
Appendix).  Relativistic boosting effects due to the pulsar wind
motion are taken into account by Lorentz transforming the particle
distribution function (Eq.~\ref{eq:FD}) from the comoving frame to the
pulsar/observer frame, see \cite{2009A&A...503...13P}.

\section{Application to \psrb}
\label{sec:Results}

In this paragraph, we first review the essential characteristics of
the \psrb\ system, then show spectral features expected from the
striped wind model before closing with some estimates of the relevant
parameters.

\subsection{Description of the binary and pulsar}

PSR B1259-63 / SS2883 is part of the small group of gamma-ray
binaries, systems sharing a number of observational properties
(notably a high level of HE or VHE gamma-ray emission compared to
X-rays; \citealt{2006A&A...456..801D}) and which have been conjectured
to all harbor rotation-powered pulsars.  Power-law spectra with
exponential cutoffs and a flux varying on the orbital period have been
observed in two other gamma-ray binaries, prompting speculations as to
whether this could be accommodated by pulsar emission (see
\S\ref{others}). PSR B1259-63 remains the only system of this group
where a pulsar has been detected.

PSR B1259-63 is a 47.76~ms radio pulsar in a 1236.7~day orbit around a
10~M$_\odot$ and 6~R$_\odot$ Be star (SS 2883) with a temperature $T_*
\approx 27 ,000$ K \citep{1994MNRAS.268..430J}. The distance $D$ to
the system is about 1.5~kpc. The temperature and distance have been
recently revised to 27,500 --34,000~K and 2.3~kpc respectively
\citep{2011arXiv1103.4636N}.  The orbit is eccentric with $e=0.87$
while the inclination of the system $i$ is about 36$\degr$, which
implies that the distance of the pulsar to the Be star varies from
9.9$\times10^{10}$~m (periastron, orbital phase $\phi_{\rm orb}=0$) to
1.4$\times 10^{12}$~m (apastron, $\phi_{\rm orb}=0.5$) and that the
angle between the Be star, pulsar and observer varies between
$\pi-i=54\degr$ (superior conjunction, $\phi_{\rm orb}\approx0.995$)
and $\pi+i= 126\degr$ (inferior conjunction, $\phi_{\rm
 orb}\approx0.045$). Hence, the stellar photon density seen by the
pulsar increases by a factor 200 at periastron, compared to apastron,
and this is also where the effects of anisotropic IC scattering should
be most manifest.  HESS observations have detected strongly
phase-dependent VHE emission around periastron, interpreted as IC
emission from pairs accelerated or randomized at the pulsar wind
termination shock \citep{2005A&A...442....1A,2009A&A...507..389A}.
The first periastron passage observable by the {\em Fermi}/LAT
occurred in mid-December 2010 and has resulted in the detection of
variable HE emission (see \S1).

The measured pulse period derivative implies an energy loss rate for
the pulsar of $\dot{E}\approx 8\ 10^{28}$~W and a magnetic field
$B\approx 3.3\ 10^{7}$~T. The radio pulsations are eclipsed close to
periastron by material in the circumstellar disk of the Be star
\citep{1995MNRAS.275..381M}. The pulse profile is reminiscent of the
Crab pulse profile, with two peaks of commensurate amplitude separated
by 145$\degr$ \citep{2002MNRAS.336.1201C}. Pulsations have not been
detected in other wavelengths; the upper limit on the amplitude in
X-rays is 2\% \citep{2006MNRAS.367.1201C}. Following
\citet{2011MNRAS.412.1870P}, radio pulsations are thought to arise
close to the polar caps, which are visible to the observer only when
$\zeta\approx \chi$. Furthermore, the double peaked radio pulse
profile in PSR~B1259-63 suggests a nearly orthogonal rotator,
constraining the obliquity $\chi\approx 90\degr$. In this case, the
average density around the pulsar is only weakly dependent on latitude
and pulsed HE emission is expected over most of the sky. The
geometrical parameters of the striped wind model are therefore
favorable and significant pulsed HE emission can be expected near
periastron due to the increased seed photon density from the Be star.

\subsection{Orbital dependence of the striped wind emission}

From the model exposed in \S\ref{sec:Model}, we compute light-curves
and spectra accounting only for photons from the companion, neglecting
all other contributions. This will give a first insight into the
orbital phase dependence of the pulse shape and phase-resolved
spectra. When discussing the results below, a clear distinction must
be kept between emission modulated by {\it orbital phase} due to the
binarity of the system and modulation induced by the {\it rotation of
the neutron star} and responsible for the pulsed emission.

As discussed above, we expect a nearly orthogonal rotator with the
line-of-sight inclination close to $90\degr$. More precisely, for the
following plots, we choose $\chi = 85\degr$ and $\zeta=55\degr$. We
take a mono-energetic photon field from the companion with an energy
of 6~eV. The wind is mildly relativistic with $\Gamma_{\rm v} = 10$.
This value for the Lorentz factor seems relatively small but is
motivated by the fact that a similar Lorentz factor ($\Gamma_{\rm v}
\approx 20$) was already used to satisfactorily fit the phase-resolved
optical polarization of the Crab pulsar \citep{2005ApJ...627L..37P}.
Moreover, in our striped wind model, pulsed emission is assumed to be
dominant very close to the light cylinder at the base of the
relativistic outflow. This fact does not exclude the possibility of a
bulk acceleration of the wind to much higher Lorentz factor, for
instance due to magnetic reconnection in the stripes
\citep{2001ApJ...547..437L}, while propagating towards the termination
shock. The particle distribution function starts from $\gamma_{\rm
  min}=10^2$ up to $\gamma_{\rm max}=10^3$ with a power-law index
$p=2$.  These values are justified in the discussion
of~\S\ref{sec:Discussion}.

\begin{figure}
\centering
\includegraphics[width=0.45\textwidth]{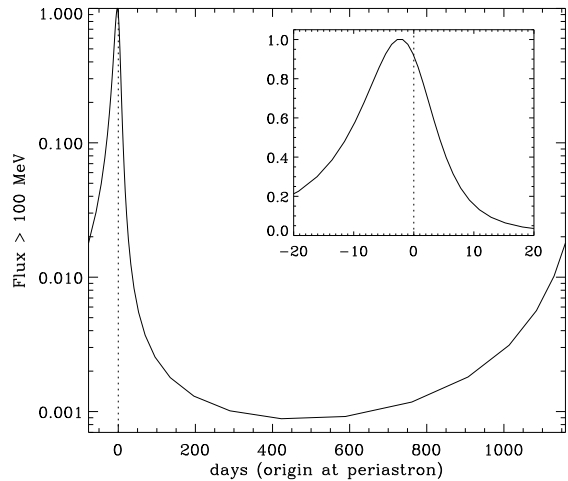}
\caption{Lightcurve vs orbital phase for the pulse averaged
  high-energy emission from the striped wind above 100~MeV. The
  inset zooms in on periastron passage.}
\label{fig:CourbeLumiereOrbitale}
\end{figure}

The orbital phase light-curve is shown in
Fig.~\ref{fig:CourbeLumiereOrbitale} for $E>100$~MeV. As expected, the
flux peaks close to periastron passage (two days before). The contrast
in flux is $\approx 700$, slightly more than the $\approx 200$
contrast that would be derived from the seed photon density because of
the angle dependence of inverse Compton scattering on a point source.
Note the asymmetrical shape with a rising time longer than the falling
time. Significant emission occurs on a short timescale, with the flux
greater than 10\% of its peak value from 28 days before periastron up
to 13 days after periastron.

\begin{figure}
\centering
\includegraphics[width=0.45\textwidth]{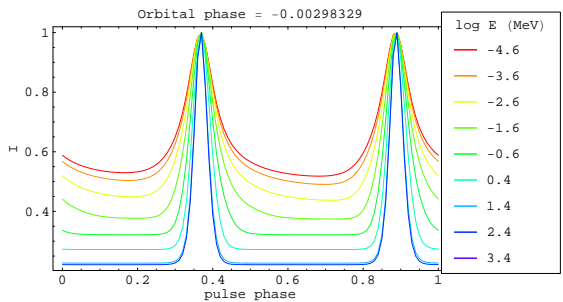}
\caption{Variation of the pulsar light-curves as a function of
  energy for a given orbital phase, here we took $\phi=-0.00298$. A
  smooth change with increasing contrast in pulses is recognizable
  when moving from lower to higher energies.  Peak intensities
    are normalized to unity.}
\label{fig:PulseEnergie}
\end{figure}

No variation of the pulse shape with orbital phase is expected because
pulses are imprints of the current sheets that are independent of the
scattering photon field, excepted for the peak intensity evolving with
orbital phase according to Fig.~\ref{fig:CourbeLumiereOrbitale}.
However, the shape is sensitive to the photon energy.  This variation
is quantified in Fig.~\ref{fig:PulseEnergie} where we plot the pulsed
light-curves against energy from X-rays ($E\approx1$~keV) to
gamma-rays ($E\approx1$~GeV, assuming electrons are accelerated up
these energies).  At lowest energies, around a few keV, a non
negligible off-pulse component remains visible between both pulses
whereas at highest energies, above several MeV, pulses sharpen with
decreasing width and vanishing off-pulse part. We emphasize that this
trend is not correlated with any orbital motion, it is an intrinsic
feature of the pulsed radiation mechanism invoked for pulsar, entirely
produced by relativistic beaming effects, let it be isolated or in
binaries.  In the striped wind scenario, if beaming disappears so does
pulsed emission. This trend in pulse variation is closely related to
the Doppler boosting of the spectral flux density from the wind frame
to the observer frame.  Indeed the Lorentz factor shapes the pulse
profiles. On one side, if the beaming remains to weak, the DC
component dominates with pulsation hardly detectable. On the other
side, for ultrarelativistic flows, the pulse profiles become
independent of the Lorentz factor and reflect the characteristics of
the stripes, i.e. width and particle density number.  This dependence
is explained as follows.  Assuming for instance a spectral slope of
index~$\alpha$ in intensity, the spectral flux density $F_\nu\propto
\nu^{-\alpha}$ will be magnified by a factor $\mathcal{D}^{2+\alpha}$
where $\mathcal{D} = 1/\Gamma_{\rm v}(1-{\bf \beta}_{\rm v} \cdot {\bf
  n})$ is the relativistic Doppler factor (see \S3.3 below). The
enhancement therefore strongly depends on this index $\alpha$ thus on
photon energy. Note that if PSR B1259-63 is an orthogonal rotator then
the time lag between gamma-ray and radio pulses should correspond to
$P/4$, $1/4$ of the pulsar period or 90$\degr$ in phase, according to
Eq.~(31) in \cite{2011MNRAS.412.1870P}.

\begin{figure}
\centering
\includegraphics[width=0.45\textwidth]{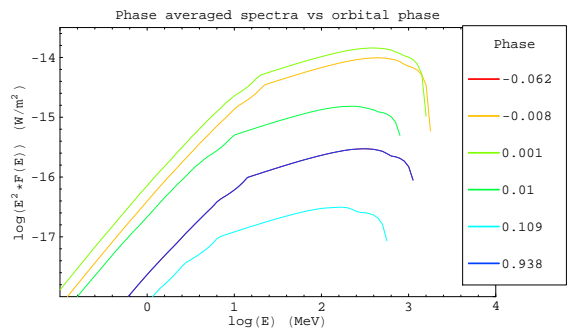}
\caption{High-energy pulse averaged spectra of the striped wind for
  different orbital phases of the binary. The physical
    parameters are estimated according to the discussion in~\S3.3.}
\label{fig:SpectreOrbitale}
\end{figure}

An example of spectra for different orbital phases is plotted in
Fig.~\ref{fig:SpectreOrbitale}. Three distinct parts corresponding to
three different slopes are visible. First, the spectrum is hard
($\alpha=0$) below the low energy cut-off around a few 10~MeV implied
by the minimum Lorentz factor of the particle power law distribution
function.  The beaming is $\propto \mathcal{D}^2$.  The flux in X-ray
(around 1~keV) is two to three orders of magnitude fainter than the
gamma-ray flux. Most of the pulsed emission is radiated around 10~MeV.
Second, between a few 10~MeV and a few GeV, the classical Thomson
regime applies with a softer spectrum $\alpha=(p-1)/2$, leading to
more effective beaming $\propto \mathcal{D}^{2+\alpha}$ and sharper
pulses.  Finally, the Klein-Nishina regime corresponding to the
softest part of the spectrum, above a few~GeV, the highest values of
$\alpha$ are reached, therefore a very effective beaming and strongly
pronounced pulses.

% A careful analysis of the orbital phase dependence reveals a change in
% the low cut-off energy, increasing when the pulsar leaves the
% neighborhood of the periastron. Meanwhile, the transition between
% Thomson and Klein-Nishina regime occurs around a few~GeV. Thus the
% spectral shape changes slightly during the pulsar trip along its
% orbit. This phenomenon should be observable if emission is detected
% over the whole orbital period.

The characteristic frequencies of the spectrum (associated with the
minimum electron energy and the Klein-Nishina transition) change
slightly with orbital phase because of the dependence of the inverse
Compton cross-section with angle. The effect would be stronger for a
more edge-on inclination of the binary orbit (see Fig. 1 in \cite{2008A&A...477..691D}).

\subsection{Estimated striped wind luminosity}

We have reviewed the basic assumptions behind the striped wind model
and we have shown that it leads to orbital modulations. Here, we give
some analytical estimates of the synchrotron and inverse Compton
luminosity from the striped wind. The relevant quantities are the
Lorentz factor~$\Gamma_{\rm v}$, the magnetic field~$B_L$ and the
particle density number~$n_L$ within the sheet at the light-cylinder
radius~$r_L$. For PSR B1259-63, $B_L$ is equal to 2.7 T at the light
cylinder radius $r_L=2.3\times 10^6$ m. $\Gamma_{\rm v}$ and $n_L$
will be constrained from the observations.

Assuming pressure balance in the current sheet between the particles
and the magnetic field implies
\begin{equation}
\label{eq:Pression}
\frac{1}{3} \, \gamma' \, n_e' \, m_e \, c^2 = \frac{B'^2}{2\,\mu_0}
\end{equation}
where the $\gamma'$ is the mean Lorentz factor of the pairs, $n_e'$ is
the particle density and $B'$ the magnetic field. Primed quantities
refer to values measured in the frame comoving with the wind and the
subscript $L$ refers to values at the light cylinder.  Introducing the
magnetization parameter defined as
\begin{equation}
\label{eq:Magnetisation}
\sigma \equiv \frac{{B'}^2}{\mu_0\,n_e'\,m_e\,c^2} = \frac{B^2}{\mu_0\,\Gamma_{\rm v}\,n_e\,m_e\,c^2}= \frac{B_L^2}{\mu_0\,\Gamma_{\rm v}\,n_{\rm L}\,m_e\,c^2}
\end{equation}
we get $\gamma' \approx 3\sigma/2$. Note that although $B'$ and $n_e'$
vary with distance to the pulsar, $\sigma$ is actually constant with
radius so that $\sigma=\sigma_L$, its value at the light cylinder
radius. Therefore, with these assumptions, $\gamma'$ is also constant
with radius.

\subsubsection{Synchrotron flux}
The synchrotron emissivity in the comoving frame, assuming isotropic radiation, is 
\begin{equation}
\label{eq:CoeffEmission}
j_{s}'(\nu') = \frac{n_e'}{4\,\pi} \,  \frac{4}{3}
  \sigma_T\,c\,\gamma'^2\,\frac{B'^2}{2\,\mu_0} \,\delta(\nu'-\nu_s')
\end{equation}
assuming that the mono-energetic particles radiate only at their synchrotron peak (cut-off) frequency 
\begin{equation}
  \label{eq:cutoff}
  \nu_s' = \frac{3}{4\,\pi} \, \gamma'^2 \, \frac{e\,B'}{m_e}=\frac{3}{4\,\pi} \, \left(\frac{3\sigma_L}{2}\right)^2 \, \frac{e\,B_L}{m_e \Gamma_{\rm v}} \left(\frac{r_L}{r}\right)\equiv \nu^\prime_{L} \left(\frac{r_L}{r}\right)
\end{equation}
where we have used that the magnetic field is purely toroidal.  The
synchrotron flux measured by a distant observer located at distance
$D$ is
\begin{equation}
\label{eq:FluxTL1}
F_{s}(\nu) = \frac{1}{D^2} \, \int j_{s}(\nu) \, dV = 
\frac{1}{D^2} \, \int \mathcal{D}^2 j'_{s}(\nu') \, dV
\end{equation}
Recall that only the stripes radiate significantly. Thus the effective
volume is much less than the total volume occupied by the wind, the
ratio between both is symbolized by~$\Delta$ which is around~0.1 and
$dV= \Delta r^2 dr d\Omega$ (note that $\Delta$ is different - but
related to - $\Delta_\varphi$).  Integrating the above using the
radial dependence of the synchrotron frequency (Eq.~\ref{eq:cutoff})
leads to
\begin{eqnarray}
\label{eq:SPectralFlux}
F_{s}(\nu) & = & \frac{2}{9 D^2} \, \sigma_T \, n_L \, B_L \, r_L^3 \,
\frac{m_e c}{\mu_0 e} \int_{\Omega_{\rm p}} \frac{\mathcal{D}^2}{\Gamma_{\rm v}^2}  \Delta d\Omega \\
&=& \frac{8\pi}{9 D^2} \, \sigma_T \, n_L \, B_L \, r_L^3 \, \frac{m_e c}{\mu_0e} \frac{f_p  \Delta}{\Gamma_{\rm v}^2} \\
%  & = & \frac{8\,\pi}{9} \, \frac{\mathcal{D}^2}{D^2} \, \sigma_T \, c \frac{B_L \, r_L^3 \, n_L \,  m_e}{\mu_0\,e\,\Gamma_{\rm v}^2} \, \Delta \\
%& \approx & \frac{32\,\pi}{9} \, \frac{1}{D^2} \, \frac{\sigma_T \, m_e \, c}{\mu_0\,e} \, B_L \, r_L^3 \, n_L \, \Delta \\
& \approx & 4 \times 10^{-45} \, \Delta \, n_L \, \Gamma_{\rm v}^{-2}\   \textrm{W m}^{-2}\, \textrm{Hz}^{-1}. 
%\left( \frac{n_L}{\textrm{m}^{-3}} \right)\  
\end{eqnarray}
where $f_p$ is a coefficient of order unity due to the integration
over the finite solid angle covered by the striped wind~$\Omega_{\rm
  p}$, usually less than $4\pi$~steradians, and which we collate with
$\Delta$.  For a more general discussion including a power-law
distribution of particle instead of a mono-energetic one, see
\cite{2002A&A...388L..29K}.

\subsubsection{Inverse Compton flux}
The inverse Compton emissivity can be approximately calculated using a
similar approach and assuming the Thomson regime (the validity of this
assumption can be verified a posteriori). If the source of seed
photons (star) is a point-like ($R_\star\ll d_\star$) blackbody of
temperature $T_\star$, the comoving radiation density of seed photons
is given by
\begin{equation}
u'_{\star}=\Gamma_{\rm v}^2 (1-\beta_v \cos \theta_\star)^2
\frac{\sigma_{\rm
  SB}T_\star^4}{c}\left(\frac{R_\star}{d_\star}\right)^2\approx 0.5
\Gamma_{\rm v}^2 \ \textrm{J\ m}^{-3}
\end{equation}
at periastron. Here, $\theta_\star$ is the angle between the direction
of the star and the direction of motion of the electron. The dominant
emission comes from those electrons that travel towards the observer.
Hence, the angle varies between 54$\degr$ and 126$\degr$ (see \S3.1)
and $(1-\beta_v \cos \theta_\star)\approx 1$. In the following, we
assume that the stellar radiation is isotropic in the observer frame.
The striped wind emission is again assumed to be monochromatic at the
frequency
\begin{equation}
 \label{eq:cutoff2}
 \nu_c' = \frac{4}{3} \, \gamma'^2 \, \nu'_\star \approx
 \frac{4}{3} \, \gamma'^2 \, \Gamma_{\rm v} \nu_\star =3\Gamma_{\rm
   v} \sigma_L^2\nu_\star
\end{equation}
where $h\nu_\star\approx 2.7 k T_\star$ is the peak frequency of the
stellar blackbody in $\nu F_\nu$. The characteristic inverse Compton
frequency is independent of radius so the high energy emission is a
line at the rest frame frequency $\nu_c=3 {\cal D} \Gamma_{\rm v}
\sigma_L^2\nu_\star\approx 35\ \Gamma_{\rm v}^2 \sigma_L^2$ eV with a
spectral flux
\begin{equation}
\label{eq:FluxTL2}
\nu_c F_{c} = \frac{1}{D^2} \iint \mathcal{D}^2\frac{n_e'}{4\,\pi} \,  \frac{4}{3}
  \sigma_T\,c\,\gamma'^2\, u'_\star \,\delta(\nu'-\nu_c') \, \Delta dV d\nu
\end{equation}
After integration, expressing the luminosity as a function of
$\Gamma_{\rm v}$ and $n_L$ gives
\begin{eqnarray}
\label{eq:FluxTL3}
\nu_c F_{c}  &\approx&\frac{3}{D^2} n_L \sigma_T c \left(\Gamma_{\rm v}\sigma_L\right)^2  u_{\star} r_L^2  l_{\rm em}\Delta\\
%&\approx&\frac{3}{D^2} n_L \sigma_T c \left(\frac{B_L^2}{\mu_0 n_L mc^2}\right)^2  u_{\rm ph} r_L^2  l_{\rm em}\Delta\\
%&\approx&\frac{3}{D^2} \frac{B_L^2}{\mu_0 mc^2} \sigma_T c \left(\Gamma_{\rm v}\sigma_L\right)  u_{\rm ph} r_L^2  L\Delta\\
&\approx& 1.2\times 10^{-20}\,  \Gamma_{\rm v}\sigma_L \, \Delta\,
\frac{l_{\rm em}}{r_L}\  \textrm{W m}^{-2} 
%&\approx& 0.9\,  n_L^{-1}\, \Delta\, \frac{l_{\rm em}}{r_L}\  \textrm{W m}^{-2} 
\end{eqnarray}
where $l_{\rm em}$ is the range of radii over which there is emission,
set in practice by particle cooling. In the above expression we have
collated with $\Delta$ the geometrical term
\begin{equation}
g_{\rm p}=\frac{1}{4\pi \Gamma_{\rm v}} \int_{\Omega_{\rm p}} \mathcal{D}^3 d\Omega\approx 1
\end{equation}
that appears in the integration and which is of order unity. Note that
$g_p=1$ only when the striped wind covers the full $4\pi$~steradians.
The inverse Compton luminosity depends on the orbital phase through
the stellar radiation density (the numerical value given here is for
periastron).

\subsubsection{Striped wind parameters in PSR B1259-63}

The expressions above are used to constrain the parameters of the
striped wind in PSR B1259-63. The spectrum of PSR B1259-63 does not
have a high enough S/N to constrain the presence of an exponential
cutoff during the first detection in December
\citep{2011arXiv1103.4108A}, when the present model is thought to
apply (see \S4). \citet{2011arXiv1103.3129T} report no detection above
1.4 GeV in the first detection and a spectral exponential cutoff
around 300 MeV during the January 2011 flare. LS 5039 and LS I
+61$\degr$303 show exponential cutoffs at a few GeV, in the range of
the HE cutoffs (or breaks) observed in pulsars. Radiative or adiabatic
losses or a complex distributions of pairs will readily widen the
mono-energetic line in the simple approximation to power-law emission
with a sharp cutoff, as observed in pulsars. We tentatively associate
the inverse Compton frequency $\nu_c$ derived above with an emission
cutoff at a few GeV, typical of the observed {\em Fermi}/LAT pulsars
\citep{2010ApJS..187..460A}, so that
\begin{equation}
h\nu_c=3 \mathcal{D} \Gamma_{\rm v} \sigma_L^2 h \nu_\star\approx 3 {\rm\ GeV.}
\end{equation}
This implies
\begin{equation}
\Gamma_{\rm v} \sigma_L=\frac{B_L^2}{\mu_0 n_L mc^2}\approx 10^4 ~~\textrm{and}~~ n_L\approx 7\times 10^{15} {\rm\ m}^{-3}.
\label{gammasig}
\end{equation}
The Lorentz factor of the electrons is $\gamma=3\sigma_L {\cal D} /2
\approx 10^4$ so inverse Compton scattering on stellar photons of
typical energy $\epsilon_\star=2.7 kT_\star\approx 6$ eV is indeed in
the Thomson regime. We are effectively setting $\gamma_{\rm max}=10^4$
(in the observer frame). As an aside, the value of $n_L$ puts a limit
on the pair multiplicity factor~$\kappa$ in the polar caps. The
Goldreich-Julian particle flux from both polar caps is
\begin{equation}
2\dot{N}_{\rm GJ}\approx4\,\pi \, \frac{\varepsilon_0\,\Omega^2 \, B_{\rm ns}
  \, R_{\rm ns}^3}{e}  \approx 4\times 10^{32}\ {\rm s}^{-1} 
\end{equation}
so the expected pair density at the light cylinder is
\begin{equation}
n_L\approx \kappa\frac{2\dot{N}_{\rm GJ}}{4\pi r_L^2 c}\approx 2\times10^{10} \, \kappa\ {\rm\ m}^{-3} 
\end{equation}
and the pair multiplicity is constrained to 
\begin{equation}
\kappa\approx 3\times 10^5 \, \Delta^{-1}.
\end{equation}
A comparable constraint for \psrb, $\kappa \lesssim 8 \times 10^4$,
was derived by \cite{2007A&A...473..683P} from considerations about
magnetic dissipation at the termination shock.

The inverse Compton luminosity at periastron directly depends on the
adopted spectral cutoff (Eq.~\ref{eq:FluxTL3}), giving
\begin{equation}
\nu_c F_{\rm c}(3\ {\rm GeV})\approx 1.2\times 10^{-16}\Delta \left(\frac{l_{\rm em}}{r_L}\right) \textrm{W m}^{-2}
\end{equation}
or $4\times 10^{24}\, \textrm{W}$, isotropic equivalent, to compare to
a pulsar spindown power of $8\times 10^{28}\, \textrm{W}$. A
significant fraction of the power is radiated in gamma rays only if a
very large range of radii ($l_{\rm em}$) are involved. At periastron,
the pulsar is at a distance from the star equivalent to $4\times10^4
r_L$. Hence, a large fraction of the pulsar power can in principle be
radiated in gamma rays if the striped wind propagates most of the way
to the star. Both particle cooling in the pulsar wind and ram pressure
from the stellar wind will act to lower the actual level of high
energy emission. The gamma-ray luminosity of PSR B1259-63 at
periastron represents about 5\% of the spindown power, so $r_{0}
\approx 1000 r_{L} \Delta^{-1}$ in order to match the observed flux.
Note that the emission from each pulse is spread over a time $l_{\rm
  em}/ \Gamma_{\rm v}^2 c$. The HE emission will remain pulsed only if
$l_{\rm em} \leq \Gamma_{\rm v}^2 r_L$ \citep{2002A&A...388L..29K},
which requires $\Gamma_{\rm v}\geq 30 \Delta^{-1/2}$.

The synchrotron spectrum is a power-law with $\nu F_{\nu}\propto \nu$.
The synchrotron frequency is no higher than its value at the light
cylinder (Eq.~\ref{eq:cutoff}) and presumably less if emission starts
away from the light cylinder. At the light cylinder,
\begin{equation}
\nu_L\approx 2.5\times 10^{11} \, \sigma_L^2 \, \textrm{Hz,}
\end{equation}
hence, the maximum synchrotron spectral luminosity is (Eq.~\ref{eq:SPectralFlux})
\begin{equation}
\nu_s F_s (r=r_L)\approx  7\times 10^{-18} \sigma_L^2 \Gamma_{\rm v}^{-2} \Delta {\rm\ W}\, \textrm{m}^{-2}
\end{equation}
This must be smaller than the spindown luminosity of the pulsar
$\approx 3\times 10^{-12} {\rm\ W}\, \textrm{m}^{-2}$ at $D=1.5$ kpc,
which yields the conservative limits $\sigma_L\leq 650\, \Gamma_{\rm
  v}\, \Delta^{-1/2}$ and $\nu_L\leq 10^{17} \, \Gamma_{\rm v}^2
\Delta^{-1} \, \textrm{Hz}$. In the above we have fixed the
characteristic gamma-ray energy to 3 GeV, imposing $\Gamma_{\rm v}
\sigma_L\approx 10^4$. Therefore, the synchrotron luminosity is no
larger than $\nu_s F_s \approx 7\times 10^{-10}\, \Gamma_{\rm v}^{-4}
\Delta {\rm\ W}\, \textrm{m}^{-2}$. The limit on spindown luminosity
gives $\Gamma_{\rm v}\geq 4\, \Delta^{1/4}$ and $h\nu_L\leq 10
\Delta^{-1/2}$ keV. A moderate Lorentz factor for the wind is enough
to significantly reduce the synchrotron luminosity. This is consistent
with the lack of a very hard component in the spectral energy
distribution and with the non-detection of X-ray pulsations. The
synchrotron flux is indeed very sensitive to the Lorentz factor of the
wind Eq.~(\ref{eq:SPectralFlux}). Much higher value than those adopted
in this work would imply an undetectable synchrotron X-ray flux.

Taking into account the distribution in energy of pairs and cooling
will spread the emission over a broader band. For instance, Thomson
cooling of a mono-energetic pair injection results in a $p=2$
distribution for the pairs and $\nu F_\nu\propto \nu^{1/2}$. Pairs
with $\gamma\approx 10^4$ have an inverse Compton cooling timescale
$\approx 600$ s at periastron so adiabatic cooling probably dominates
over radiative cooling. Adiabatic cooling of mono-energetic pairs
gives a $p=1$ distribution and $\nu F_\nu\propto \nu$. The spectrum
also has this slope below $\gamma_{\rm min}$ if the injection is a
power-law (Fig. 4). The pulsed inverse Compton spectral luminosity in
X-rays at 3 keV is thus expected to be from $10^3$ to $10^6$ times
lower than at 3 GeV, consistent with the current upper limit on the
X-ray pulsed fraction, 2\% of $2\times 10^{-14}$ W\, m$^{-2}$
\citep{2006MNRAS.367.1201C}.

\section{Discussion}
\label{sec:Discussion}

The striped wind model in the environment of a massive star naturally
leads to orbitally-modulated inverse Compton emission in the GeV
range. The model involves several parameters, some attributed to the
pulsar wind, like $(\chi, \Gamma_{\rm v},N,N_0,\Delta_\varphi)$,
supplemented with a prescribed particle distribution function, and
some attributed to the binarity of the system, like star temperature,
radius, orbit inclination angle, eccentricity, periastron position and
photon density number, respectively $(T_\star, R_\star,e,\omega,n_{\rm
 ph})$. As shown in the previous section, the pulsar wind parameters
influence the observability of high energy emission, since pulsed
emission from the striped wind can be detected only if the
line-of-sight crosses the stripes, and the shape of this pulsed
emission \citep{2011MNRAS.412.1870P}. The binary parameters influence
the shape of the orbital modulation \citep{2008A&A...477..691D}.

\subsection{{\em Fermi}/LAT detections of PSR B1259-63}

The orbital parameters are relatively well known in PSR B1259-63.
Pulsed emission from the striped wind can be detected only if the
line-of-sight crosses the stripes. PSR~B1259-63 is close to an
orthogonal rotator owing to its double radio-pulse structure separated
by roughly half a period (possible only if both magnetic poles are
visible). As a consequence, stripes should fill the whole sky and high
energy emission can be detected. Maximum inverse Compton flux is
expected near periastron. We showed that, assuming the high energy
emission cuts off at a few GeV like in pulsars or other gamma-ray
binaries, the level of synchrotron and inverse Compton emission can be
derived and is consistent with observations at periastron. We propose
that the initial detection of gamma-ray emission from PSR B1259-63 is
due to emission from the striped pulsar wind.

Curiously, much stronger HE emission was reported about a month after
periastron passage. The reported flux level $\approx 1.5 \times
10^{-6}\, \textrm{ph cm}^{-2} \textrm{s}^{-1}$ is 10 times higher than
during the first passage, corresponding to a large fraction of the
spindown luminosity
\citep{2011arXiv1103.4108A,2011arXiv1103.3129T}. Although a high
gamma-ray luminosity is possible (\S3.3.3), this is not expected to
occur after periastron passage since both the stellar radiation
density is decreasing with distance and the geometry for inverse
Compton scattering with stellar photons becomes less favorable
(increasing interaction angle).  The increase could be unrelated to
the striped wind. If it is, additional contributions to the seed
photon field could be involved.

For instance, the surrounding pulsar wind nebula (PWN) beyond the
termination shock can contribute to the source of target photons. In
binaries, the termination shock has a cone-like appearance, pointing
more or less away from the star \citep{2010A&A...516A..18D}, whose
opening angle depends on the ratio of momentum fluxes from the stellar
and pulsar winds \citep{2008MNRAS.387...63B}. The PWN emits mostly
X-rays and gamma-rays with a luminosity bounded by the spindown
luminosity, which is about 100 times lower than the luminosity of the
Be star. The termination shock is typically expected to be closer than
the star by a factor 10 (e.g. \citealt{1997ApJ...477..439T};
\citealt{2006A&A...456..801D}), compensating the luminosity, but the
interactions are less efficient as they occur in the Klein-Nishina
regime. However, after periastron, the orbital geometry is such that
the pulsar gradually moves in between the star and the
observer. Hence, the cone moves in between the pulsar and the
observer. If the cone has a large opening angle, pairs moving along
the line-of-sight then encounter PWN photons with a progressively
smaller interaction angle (closer to head on) than at periastron,
while the angle with stellar photons increases (from 54$\degr$ at
superior conjunction, a week before periastron passage, to 126$\degr$
at inferior conjunction, 55 days after periastron passage). As with
external Compton emission in the relativistic jets of AGNs, the
emission will be highly anisotropic. Doppler boosting of the PWN
radiation density seen by the pairs moving towards the observer (in
the direction of the PWN) and the increased upscattering rate along
the line-of-sight can result in a much stronger gamma-ray flux. The
steeper spectrum reported for the second detection could be due to the
scattering occurring in the Klein-Nishina regime.

For completeness, we also note that emission from the Be disk
\citep{2011MNRAS.tmp...17V} or thermal emission from the neutron star
could also contribute significantly to the seed photon radiation
density (although the scattering geometry is unfavorable in the latter
case). We defer to future work an investigation of these processes and
their relevance to explain the HE gamma-ray lightcurve after
periastron passage.

\subsection{Evidence for a striped wind in other gamma-ray binaries\label{others}}

Two binaries, LS~5039 and LS~I +61$\degr$303, are established HE
gamma-ray sources \citep{2009ApJ...701L.123A,2009ApJ...706L..56A} and
have been conjectured to host pulsars with a large spindown power like
PSR B1259-63 \citep{2006A&A...456..801D}.  {\em Fermi}/LAT
observations showed that the HE gamma-ray emission does not connect
naturally to the VHE emission detected by ground based Cherenkov
arrays above 100 GeV. In both cases, there is an exponential cutoff at
a few GeV, typical of the HE spectra of a gamma-ray pulsar. The HE
emission appears to be due to a different population of particles than
in VHE regime because of this cutoff. A straightforward solution is to
interpret the HE as emission coming from the vicinity of the pulsar
and the radio/X-ray/VHE as PWN emission arising beyond the termination
shock.

{\em Fermi}/LAT observations showed that the HE gamma-ray flux from
LS~5039 and LS I +61$\degr$303 is modulated on the orbital period,
which would be difficult to explain via the usual magnetospheric
emission processes \citep{2009ApJ...706L..56A} but is readily
explained by the striped wind model developed in the previous
sections. The HE gamma-ray emission should be pulsed but this is
difficult to verify in a binary without an ephemeris for the
pulsation.

The conditions are very similar to PSR B1259-63 except for the smaller
orbits. The radiation density at periastron is greater by a factor
$\approx 50-1000$ and this implies a larger high gamma-ray
luminosity. Indeed, the mean flux above 100 MeV is $\approx 10^{-6}\,
\textrm{ph cm}^{-2}\, \textrm{s}^{-1}$. The orbital modulations are
consistent with expectations from anisotropic inverse Compton
scattering \citep{2009ApJ...701L.123A,2009ApJ...706L..56A}, which also
apply to the striped wind model as discussed in the previous sections.

Interestingly, the HE lightcurve of LS I +61$\degr$303 peaks slightly
later than expected, like PSR B1259-63. As discussed above, this might
be explained if there is an additional source of seed photons besides
the star, notably the PWN. Changes in the source of seed photons would
also need to be invoked to explain the long term or orbit-to-orbit
variability in the GeV lightcurve of LS I +61$\degr$303. Long term
variability could also be attributed to pulsar precession. Such free
precession has indeed been observed in pulsars like for PSR~B1828-11
\citep{2000Natur.406..484S} for which timescales of several months
have been reported. Precession implies changes in the orientation of
the striped wind with respect to the orbital plane, inducing a change
in the particle density number along the line of sight.

A favorable orientation of the pulsar is required to invoke striped
wind emission, implying the number of gamma-ray binaries may actually
be greater than currently detected. If the distribution of obliquities
is isotropic then a random line-of-sight crosses the striped wind in
half of the pulsars. The model predicts a population of sources with
strong PWN emission (less susceptible to orientation issues) but weak
or absent HE gamma-ray emission. Such systems would be most easily
detected through their VHE gamma-ray emission by ground-based
Cherenkov arrays. An example could be HESS J0632+057, a system very
strongly suspected to be a gamma-ray binary
\citep{2009ApJ...690L.101H} but that has not been detected yet by the
{\em Fermi}/LAT.

Finally, another important issue is the exact composition of the wind.
Whereas a supply of electron/positron is easily achieved by pair
creation in the polar caps, yielding a high multiplicity factor, the
only possible source for ions would be from the stellar crust itself.
We emphasize that despite the much lower density of ions, they can
drastically affect the dynamics of the striped wind because of the
large separation of time and spatial scales induced by the large mass
ratio between leptons and ions.  Indeed, the growth rates of some
instabilities can significantly deviate from a pure electron/positron
plasma and perturb the dissipation of magnetic field lines within the
stripes, modifying the efficiency of the bulk acceleration of the
wind.  However, there is so far no clear evidence for such an ion
component. The possible presence of ions in the striped wind has to be
distinguished from the hadronic interpretation of the TeV light-curves
expected from the {\it shocked wind}. In the latter, interaction of
the pulsar wind with the companion star disk material could lead to
significant emission but this emission would be unpulsed.

\section{Conclusion}

We have investigated the implications of the striped pulsar wind model
in the context of gamma-ray binaries. Inverse Compton upscattering of
stellar photons from the companion by pairs in the stripes generates
high energy gamma-ray radiation that can be detected if the
line-of-sight crosses the striped region of the wind (\S2). The
gamma-ray emission is expected to be pulsed and is also modulated
along the orbit because of the changing photon density and scattering
interaction angle (\S3).

The weak HE gamma-ray emission from PSR B1259-63 detected near
periastron can be explained by striped wind emission. The radio
pulsations suggest PSR B1259-63 is an orthogonal rotator so that the
line-of-sight always crosses the stripes. The inverse Compton emission
is maximum close to periastron. The energy of the pairs is constant
with distance from the pulsar if pressure balance between pairs and
magnetic field is assumed in the stripes and cooling is neglected.
Hence, a characteristic inverse Compton energy is expected that we
associate with a spectral cutoff at a few GeV, by analogy with the
spectra observed in the gamma-ray binaries LS 5039 and LS I
+61$\degr$303. The level of synchrotron and inverse Compton flux
depends on the wind Lorenz factor $\Gamma_{\rm v}$, the magnetization
$\sigma_L$ and the volume occupied by the stripes.  We find that a few
\% of the spindown luminosity can be radiated in gamma-rays if the
emission occurs over a large range of radii (\S3).

The gamma-ray emission should show a double-peaked pulsation at the
rotation period of the neutron star with the pulse shape providing
further diagnostics. \citet{2011arXiv1103.4108A} searched for the
pulsation but did not detect it (although, as they point out, PSR
B1259-63 is well within the range of $\dot{E}$ and $B_L$ where pulsed
gamma-ray emission is typically detected from pulsars by the {\em
  Fermi}/LAT). Detecting the gamma-ray pulsation in this binary may be
difficult to achieve because of the weak flux and variations in timing
solution around periastron \citep{2004MNRAS.351..599W}.  Pulsations
may also be smeared out if the range of emitting radii is large and
the Lorentz factor of the wind is low (\S3).

The second, much stronger HE gamma-ray detection of PSR B1259-63 that
occurred about a month after periastron passage is not explained in
this model without additional assumptions. One possibility is
additional seed photons provided by the PWN: in this case, the peak
contribution arises after periastron when the PWN is in between the
observer and the pulsar, a favorable configuration for inverse Compton
scattering (\S4.1). We note that a similar shift in phase of peak HE
gamma-ray emission is observed in LS I +61$\degr$303 and could be
explained similarly.

Emission from a striped pulsar wind might solve the puzzle of the
pulsar-like HE gamma-ray spectrum with a flux modulated on the orbital
period in the gamma-ray binaries LS 5039 and LS I +61$\degr$303. The
HE emission is stronger than in PSR B1259-63 because of the higher
stellar photon densities. 
%Pulsed HE emission is expected, although
%this might be smeared if the emitting region has a large radial extent. 
Significant HE gamma-ray emission requires that the stripes
are oriented towards us. Systems where this is not the case could
still be detected through their radio, X-ray or VHE emission
(attributed to the PWN) but will not display strong HE emission. One
such system might be HESS J0632+057 (\S4.2).

Gamma-ray binaries may thus be providing indirect evidence for striped
wind emission. If verified by the detection of pulsations in HE gamma
rays, this would bolster the view that pulsar HE emission originates
beyond the light cylinder.

\section*{Acknowledgments}

We thank D. Khangulyan and V. Bosch-Ramon for pointing out an error in
the first draft of this paper, B. Cerutti and G. Henri for helpful
discussions. This work was supported by the European Community via
contract ERC-StG-200911.

%\bibliographystyle{/home/petri/texmf/tex/latex/aa-package/bibtex/aa}%mn2e}
%\bibliography{/data/petri/habilitation/bibliotot,/data/petri/habilitation/publi}

\begin{thebibliography}{38}
 \expandafter\ifx\csname natexlab\endcsname\relax\def\natexlab#1{#1}\fi

\bibitem[{{Abdo} {et~al.}(2010{\natexlab{a}}){Abdo}, {Ackermann}, {Ajello},
 {Atwood}, {Axelsson}, {Baldini}, {Ballet}, {Barbiellini}, {Baring},
 {Bastieri}, {Baughman}, {Bechtol}, {Bellazzini}, {Berenji}, {Blandford},
 {Bloom}, {Bonamente}, {Borgland}, {Bregeon}, {Brez}, {Brigida}, {Bruel},
 {Burnett}, {Buson}, {Caliandro}, {Cameron}, {Camilo}, {Caraveo},
 {Casandjian}, {Cecchi}, {{\c C}elik}, {Charles}, {Chekhtman}, {Cheung},
 {Chiang}, {Ciprini}, {Claus}, {Cognard}, {Cohen-Tanugi}, {Cominsky},
 {Conrad}, {Corbet}, {Cutini}, {den Hartog}, {Dermer}, {de Angelis}, {de
 Luca}, {de Palma}, {Digel}, {Dormody}, {Silva}, {Drell}, {Dubois}, {Dumora},
 {Espinoza}, {Farnier}, {Favuzzi}, {Fegan}, {Ferrara}, {Focke}, {Fortin},
 {Frailis}, {Freire}, {Fukazawa}, {Funk}, {Fusco}, {Gargano}, {Gasparrini},
 {Gehrels}, {Germani}, {Giavitto}, {Giebels}, {Giglietto}, {Giommi},
 {Giordano}, {Glanzman}, {Godfrey}, {Gotthelf}, {Grenier}, {Grondin}, {Grove},
 {Guillemot}, {Guiriec}, {Gwon}, {Hanabata}, {Harding}, {Hayashida}, {Hays},
 {Hughes}, {Jackson}, {J{\'o}hannesson}, {Johnson}, {Johnson}, {Johnson},
 {Johnson}, {Johnston}, {Kamae}, {Kanbach}, {Kaspi}, {Katagiri}, {Kataoka},
 {Kawai}, {Kerr}, {Kn{\"o}dlseder}, {Kocian}, {Kramer}, {Kuss}, {Lande},
 {Latronico}, {Lemoine-Goumard}, {Livingstone}, {Longo}, {Loparco}, {Lott},
 {Lovellette}, {Lubrano}, {Lyne}, {Madejski}, {Makeev}, {Manchester},
 {Marelli}, {Mazziotta}, {McConville}, {McEnery}, {McGlynn}, {Meurer},
 {Michelson}, {Mineo}, {Mitthumsiri}, {Mizuno}, {Moiseev}, {Monte}, {Monzani},
 {Morselli}, {Moskalenko}, {Murgia}, {Nakamori}, {Nolan}, {Norris}, {Noutsos},
 {Nuss}, {Ohsugi}, {Omodei}, {Orlando}, {Ormes}, {Ozaki}, {Paneque},
 {Panetta}, {Parent}, {Pelassa}, {Pepe}, {Pesce-Rollins}, {Piron}, {Porter},
 {Rain{\`o}}, {Rando}, {Ransom}, {Ray}, {Razzano}, {Rea}, {Reimer}, {Reimer},
 {Reposeur}, {Ritz}, {Rodriguez}, {Romani}, {Roth}, {Ryde}, {Sadrozinski},
 {Sanchez}, {Sander}, {Saz Parkinson}, {Scargle}, {Schalk}, {Sellerholm},
 {Sgr{\`o}}, {Siskind}, {Smith}, {Smith}, {Spandre}, {Spinelli}, {Stappers},
 {Starck}, {Striani}, {Strickman}, {Strong}, {Suson}, {Tajima}, {Takahashi},
 {Takahashi}, {Tanaka}, {Thayer}, {Thayer}, {Theureau}, {Thompson},
 {Thorsett}, {Tibaldo}, {Tibolla}, {Torres}, {Tosti}, {Tramacere}, {Uchiyama},
 {Usher}, {Van Etten}, {Vasileiou}, {Venter}, {Vilchez}, {Vitale}, {Waite},
 {Wang}, {Wang}, {Watters}, {Weltevrede}, {Winer}, {Wood}, {Ylinen}, \&
 {Ziegler}}]{2010ApJS..187..460A}
{Abdo}, A.~A., {Ackermann}, M., {Ajello}, M., {et~al.} 2010{\natexlab{a}},
 \apjs, 187, 460

\bibitem[{{Abdo} {et~al.}(2009{\natexlab{a}}){Abdo}, {Ackermann}, {Ajello},
 {Atwood}, {Axelsson}, {Baldini}, {Ballet}, {Barbiellini}, {Bastieri},
 {Baughman}, {Bechtol}, {Bellazzini}, {Berenji}, {Blandford}, {Bloom},
 {Bonamente}, {Borgland}, {Bregeon}, {Brez}, {Brigida}, {Bruel}, {Burnett},
 {Caliandro}, {Cameron}, {Caraveo}, {Casandjian}, {Cavazzuti}, {Cecchi}, {{\c
 C}elik}, {Charles}, {Chaty}, {Chekhtman}, {Cheung}, {Chiang}, {Ciprini},
 {Claus}, {Cohen-Tanugi}, {Cominsky}, {Conrad}, {Corbel}, {Corbet}, {Cutini},
 {Dermer}, {de Angelis}, {de Luca}, {de Palma}, {Digel}, {Dormody}, {do Couto
 e Silva}, {Drell}, {Dubois}, {Dubus}, {Dumora}, {Farnier}, {Favuzzi},
 {Fegan}, {Focke}, {Frailis}, {Fukazawa}, {Funk}, {Fusco}, {Gargano},
 {Gasparrini}, {Gehrels}, {Germani}, {Giebels}, {Giglietto}, {Giordano},
 {Glanzman}, {Godfrey}, {Grenier}, {Grondin}, {Grove}, {Guillemot}, {Guiriec},
 {Hanabata}, {Harding}, {Hayashida}, {Hays}, {Hill}, {Hughes},
 {J{\'o}hannesson}, {Johnson}, {Johnson}, {Johnson}, {Johnson}, {Kamae},
 {Katagiri}, {Kataoka}, {Kawai}, {Kerr}, {Kn{\"o}dlseder}, {Kocian}, {Kuehn},
 {Kuss}, {Lande}, {Larsson}, {Latronico}, {Longo}, {Loparco}, {Lott},
 {Lovellette}, {Lubrano}, {Madejski}, {Makeev}, {Marelli}, {Mazziotta},
 {McEnery}, {Meurer}, {Michelson}, {Mitthumsiri}, {Mizuno}, {Monte},
 {Monzani}, {Morselli}, {Moskalenko}, {Murgia}, {Nolan}, {Nuss}, {Ohsugi},
 {Okumura}, {Omodei}, {Orlando}, {Ormes}, {Paneque}, {Panetta}, {Parent},
 {Pelassa}, {Pepe}, {Pesce-Rollins}, {Piron}, {Porter}, {Rain{\`o}}, {Rando},
 {Ray}, {Razzano}, {Rea}, {Reimer}, {Reimer}, {Reposeur}, {Ritz}, {Rochester},
 {Rodriguez}, {Romani}, {Ryde}, {Sadrozinski}, {Sanchez}, {Sander}, {Saz
 Parkinson}, {Scargle}, {Sgr{\`o}}, {Shaw}, {Sierpowska-Bartosik}, {Siskind},
 {Smith}, {Smith}, {Spandre}, {Spinelli}, {Striani}, {Strickman}, {Suson},
 {Tajima}, {Takahashi}, {Takahashi}, {Tanaka}, {Thayer}, {Thayer}, {Thompson},
 {Tibaldo}, {Torres}, {Tosti}, {Tramacere}, {Uchiyama}, {Usher}, {Vasileiou},
 {Vilchez}, {Vitale}, {Waite}, {Wang}, {Winer}, {Wood}, {Ylinen}, \&
 {Ziegler}}]{2009ApJ...701L.123A}
{Abdo}, A.~A., {Ackermann}, M., {Ajello}, M., {et~al.} 2009{\natexlab{a}},
 \apjl, 701, L123

\bibitem[{{Abdo} {et~al.}(2009{\natexlab{b}}){Abdo}, {Ackermann}, {Ajello},
 {Atwood}, {Axelsson}, {Baldini}, {Ballet}, {Barbiellini}, {Bastieri},
 {Baughman}, {Bechtol}, {Bellazzini}, {Berenji}, {Blandford}, {Bloom},
 {Bonamente}, {Borgland}, {Bregeon}, {Brez}, {Brigida}, {Bruel}, {Burnett},
 {Buson}, {Caliandro}, {Cameron}, {Caraveo}, {Casandjian}, {Cavazzuti},
 {Cecchi}, {{\c C}elik}, {Chaty}, {Chekhtman}, {Cheung}, {Chiang}, {Ciprini},
 {Claus}, {Cohen-Tanugi}, {Cominsky}, {Conrad}, {Corbel}, {Corbet}, {Cutini},
 {Dermer}, {de Angelis}, {de Palma}, {Digel}, {Silva}, {Drell}, {Dubois},
 {Dubus}, {Dumora}, {Farnier}, {Favuzzi}, {Fegan}, {Focke}, {Fortin},
 {Frailis}, {Fukazawa}, {Funk}, {Fusco}, {Gargano}, {Gasparrini}, {Gehrels},
 {Germani}, {Giebels}, {Giglietto}, {Giordano}, {Glanzman}, {Godfrey},
 {Grenier}, {Grondin}, {Grove}, {Guillemot}, {Guiriec}, {Hanabata}, {Harding},
 {Hayashida}, {Hays}, {Hill}, {Horan}, {Hughes}, {Jackson}, {J{\'o}hannesson},
 {Johnson}, {Johnson}, {Johnson}, {Kamae}, {Katagiri}, {Kataoka}, {Kawai},
 {Kerr}, {Kn{\"o}dlseder}, {Kocian}, {Kuehn}, {Kuss}, {Lande}, {Larsson},
 {Latronico}, {Lemoine-Goumard}, {Longo}, {Loparco}, {Lott}, {Lovellette},
 {Lubrano}, {Madejski}, {Makeev}, {Marelli}, {Mazziotta}, {McEnery}, {Meurer},
 {Michelson}, {Mitthumsiri}, {Mizuno}, {Moiseev}, {Monte}, {Monzani},
 {Morselli}, {Moskalenko}, {Murgia}, {Nolan}, {Norris}, {Nuss}, {Ohsugi},
 {Omodei}, {Orlando}, {Ormes}, {Ozaki}, {Paneque}, {Panetta}, {Parent},
 {Pelassa}, {Pepe}, {Pesce-Rollins}, {Piron}, {Porter}, {Rain{\`o}}, {Rando},
 {Ray}, {Razzano}, {Rea}, {Reimer}, {Reimer}, {Reposeur}, {Ritz}, {Rochester},
 {Rodriguez}, {Romani}, {Roth}, {Ryde}, {Sadrozinski}, {Sanchez}, {Sander},
 {Saz Parkinson}, {Scargle}, {Sgr{\`o}}, {Sierpowska-Bartosik}, {Siskind},
 {Smith}, {Smith}, {Spandre}, {Spinelli}, {Strickman}, {Suson}, {Tajima},
 {Takahashi}, {Takahashi}, {Tanaka}, {Tanaka}, {Thayer}, {Thompson},
 {Tibaldo}, {Torres}, {Tosti}, {Tramacere}, {Uchiyama}, {Usher}, {Vasileiou},
 {Venter}, {Vilchez}, {Vitale}, {Waite}, {Wallace}, {Wang}, {Winer}, {Wood},
 {Ylinen}, \& {Ziegler}}]{2009ApJ...706L..56A}
{Abdo}, A.~A., {Ackermann}, M., {Ajello}, M., {et~al.} 2009{\natexlab{b}},
 \apjl, 706, L56

\bibitem[{{Abdo} {et~al.}(2011{\natexlab{a}}){Abdo}, {Fermi LAT Collaboration},
 {Chernyakova}, {Neronov}, {Roberts}, \& {Fermi Pulsar Timing
 Consortium}}]{2011arXiv1103.4108A}
{Abdo}, A.~A., {Fermi LAT Collaboration}, {Chernyakova}, M., {et~al.}
 2011{\natexlab{a}}, ArXiv e-prints 1103.4108

\bibitem[{{Abdo} {et~al.}(2011{\natexlab{b}}){Abdo}, {Parent}, {Dubois}, \&
{Roberts}}]{2011ATel.3115....1A}
{Abdo}, A.~A., {Parent}, D., {Dubois}, R., \& {Roberts}, M. 2011{\natexlab{b}},
The Astronomer's Telegram, 3115, 1

\bibitem[{{Abdo} {et~al.}(2010{\natexlab{b}}){Abdo}, {Parent}, {Grove},
{Caliandro}, {Roberts}, {Johnston}, \& {Chernyakova}}]{2010ATel.3085....1A}
{Abdo}, A.~A., {Parent}, D., {Grove}, J.~E., {et~al.} 2010{\natexlab{b}}, The
Astronomer's Telegram, 3085, 1

\bibitem[{{Aharonian} {et~al.}(2009){Aharonian}, {Akhperjanian}, {Anton},
 {Barres de Almeida}, {Bazer-Bachi}, {Becherini}, {Behera}, {Bernl{\"o}hr},
 {Bochow}, {Boisson}, {Bolmont}, {Borrel}, {Brucker}, {Brun}, {Brun},
 {B{\"u}hler}, {Bulik}, {B{\"u}sching}, {Boutelier}, {Chadwick},
 {Charbonnier}, {Chaves}, {Cheesebrough}, {Chounet}, {Clapson}, {Coignet},
 {Dalton}, {Daniel}, {Davids}, {Degrange}, {Deil}, {Dickinson},
 {Djannati-Ata{\"i}}, {Domainko}, {O'C.~Drury}, {Dubois}, {Dubus}, {Dyks},
 {Dyrda}, {Egberts}, {Emmanoulopoulos}, {Espigat}, {Farnier}, {Feinstein},
 {Fiasson}, {F{\"o}rster}, {Fontaine}, {F{\"u}{\ss}ling}, {Gabici}, {Gallant},
 {G{\'e}rard}, {Gerbig}, {Giebels}, {Glicenstein}, {Gl{\"u}ck}, {Goret},
 {G{\"o}ring}, {Hauser}, {Hauser}, {Heinz}, {Heinzelmann}, {Henri}, {Hermann},
 {Hinton}, {Hoffmann}, {Hofmann}, {Holleran}, {Hoppe}, {Horns},
 {Jacholkowska}, {de Jager}, {Jahn}, {Jung}, {Katarzy{\'n}ski}, {Katz},
 {Kaufmann}, {Kerschhaggl}, {Khangulyan}, {Kh{\'e}lifi}, {Keogh}, {Klochkov},
 {Klu{\'z}niak}, {Kneiske}, {Komin}, {Kosack}, {Kossakowski}, {Lamanna},
 {Lenain}, {Lohse}, {Marandon}, {Martineau-Huynh}, {Marcowith}, {Masbou},
 {Maurin}, {McComb}, {Medina}, {Moderski}, {Moulin}, {Naumann-Godo}, {de
 Naurois}, {Nedbal}, {Nekrassov}, {Nicholas}, {Niemiec}, {Nolan}, {Ohm},
 {Olive}, {de O{\~n}a Wilhelmi}, {Orford}, {Ostrowski}, {Panter}, {Paz
 Arribas}, {Pedaletti}, {Pelletier}, {Petrucci}, {Pita}, {P{\"u}hlhofer},
 {Punch}, {Quirrenbach}, {Raubenheimer}, {Raue}, {Rayner}, {Renaud}, {Rieger},
 {Ripken}, {Rob}, {Rosier-Lees}, {Rowell}, {Rudak}, {Rulten}, {Ruppel},
 {Sahakian}, {Santangelo}, {Schlickeiser}, {Sch{\"o}ck}, {Schwanke},
 {Schwarzburg}, {Schwemmer}, {Shalchi}, {Sikora}, {Skilton}, {Sol},
 {Spangler}, {Stawarz}, {Steenkamp}, {Stegmann}, {Stinzing}, {Superina},
 {Szostek}, {Tam}, {Tavernet}, {Terrier}, {Tibolla}, {Tluczykont}, {van
 Eldik}, {Vasileiadis}, {Venter}, {Venter}, {Vialle}, {Vincent}, {Vivier},
 {V{\"o}lk}, {Volpe}, {Wagner}, {Ward}, {Zdziarski}, \&
 {Zech}}]{2009A&A...507..389A}
{Aharonian}, F., {Akhperjanian}, A.~G., {Anton}, G., {et~al.} 2009, \aap, 507,
 389

\bibitem[{{Aharonian} {et~al.}(2005){Aharonian}, {Akhperjanian}, {Aye},
 {Bazer-Bachi}, {Beilicke}, {Benbow}, {Berge}, {Berghaus}, {Bernl{\"o}hr},
 {Boisson}, {Bolz}, {Braun}, {Breitling}, {Brown}, {Bussons Gordo},
 {Chadwick}, {Chounet}, {Cornils}, {Costamante}, {Degrange},
 {Djannati-Ata{\"i}}, {O'C.~Drury}, {Dubus}, {Emmanoulopoulos}, {Espigat},
 {Feinstein}, {Fleury}, {Fontaine}, {Fuchs}, {Funk}, {Gallant}, {Giebels},
 {Gillessen}, {Glicenstein}, {Goret}, {Hadjichristidis}, {Hauser},
 {Heinzelmann}, {Henri}, {Hermann}, {Hinton}, {Hofmann}, {Holleran}, {Horns},
 {de Jager}, {Johnston}, {Kh{\'e}lifi}, {Kirk}, {Komin}, {Konopelko},
 {Latham}, {Le Gallou}, {Lemi{\`e}re}, {Lemoine-Goumard}, {Leroy},
 {Martineau-Huynh}, {Lohse}, {Marcowith}, {Masterson}, {McComb}, {de Naurois},
 {Nolan}, {Noutsos}, {Orford}, {Osborne}, {Ouchrif}, {Panter}, {Pelletier},
 {Pita}, {P{\"u}hlhofer}, {Punch}, {Raubenheimer}, {Raue}, {Raux}, {Rayner},
 {Redondo}, {Reimer}, {Reimer}, {Ripken}, {Rob}, {Rolland}, {Rowell},
 {Sahakian}, {Saug{\'e}}, {Schlenker}, {Schlickeiser}, {Schuster}, {Schwanke},
 {Siewert}, {Skj{\ae}raasen}, {Sol}, {Steenkamp}, {Stegmann}, {Tavernet},
 {Terrier}, {Th{\'e}oret}, {Tluczykont}, {Vasileiadis}, {Venter}, {Vincent},
 {V{\"o}lk}, \& {Wagner}}]{2005A&A...442....1A}
{Aharonian}, F., {Akhperjanian}, A.~G., {Aye}, K., {et~al.} 2005, \aap, 442, 1

\bibitem[{{Bai} \& {Spitkovsky}(2010)}]{2010ApJ...715.1282B}
{Bai}, X. \& {Spitkovsky}, A. 2010, \apj, 715, 1282

\bibitem[{{Ball} \& {Kirk}(2000)}]{2000APh....12..335B}
{Ball}, L. \& {Kirk}, J.~G. 2000, Astroparticle Physics, 12, 335

\bibitem[Bogovalov et al.(2008)]{2008MNRAS.387...63B} Bogovalov, S.~V., 
Khangulyan, D.~V., Koldoba, A.~V., Ustyugova, G.~V., 
\& Aharonian, F.~A.\ 2008, \mnras, 387, 63 

\bibitem[{{Bogovalov}(1999)}]{1999A&A...349.1017B}
{Bogovalov}, S.~V. 1999, \aap, 349, 1017

\bibitem[{{Cerutti} {et~al.}(2008){Cerutti}, {Dubus}, \&
 {Henri}}]{2008A&A...488...37C}
{Cerutti}, B., {Dubus}, G., \& {Henri}, G. 2008, \aap, 488, 37

\bibitem[{{Cheng}(2009)}]{2009ASSL..357..481C}
{Cheng}, K.~S. 2009, in Astrophysics and Space Science Library, Vol. 357,
 Astrophysics and Space Science Library, ed. {W.~Becker}, 481--+

\bibitem[{{Chernyakova} {et~al.}(2006){Chernyakova}, {Neronov}, {Lutovinov},
 {Rodriguez}, \& {Johnston}}]{2006MNRAS.367.1201C}
{Chernyakova}, M., {Neronov}, A., {Lutovinov}, A., {Rodriguez}, J., \&
 {Johnston}, S. 2006, \mnras, 367, 1201

\bibitem[{{Connors} {et~al.}(2002){Connors}, {Johnston}, {Manchester}, \&
 {McConnell}}]{2002MNRAS.336.1201C}
{Connors}, T.~W., {Johnston}, S., {Manchester}, R.~N., \& {McConnell}, D. 2002,
 \mnras, 336, 1201

\bibitem[{{Coroniti}(1990)}]{1990ApJ...349..538C}
{Coroniti}, F.~V. 1990, \apj, 349, 538

\bibitem[{{Dubus}(2006)}]{2006A&A...456..801D}
{Dubus}, G. 2006, \aap, 456, 801

\bibitem[{{Dubus} {et~al.}(2008){Dubus}, {Cerutti}, \&
 {Henri}}]{2008A&A...477..691D}
{Dubus}, G., {Cerutti}, B., \& {Henri}, G. 2008, \aap, 477, 691

\bibitem[{{Dubus} {et~al.}(2010){Dubus}, {Cerutti}, \&
 {Henri}}]{2010A&A...516A..18D}
{Dubus}, G., {Cerutti}, B., \& {Henri}, G. 2010, \aap, 516, A18+

\bibitem[{{Harding}(2009)}]{2009ASSL..357..521H}
{Harding}, A.~K. 2009, in Astrophysics and Space Science Library, Vol. 357,
 Astrophysics and Space Science Library, ed. {W.~Becker}, 521--+

\bibitem[{{Hinton} {et~al.}(2009){Hinton}, {Skilton}, {Funk}, {Brucker},
  {Aharonian}, {Dubus}, {Fiasson}, {Gallant}, {Hofmann}, {Marcowith}, \&
  {Reimer}}]{2009ApJ...690L.101H}
{Hinton}, J.~A., {Skilton}, J.~L., {Funk}, S., {et~al.} 2009, \apjl, 690, L101

\bibitem[{{Johnston} {et~al.}(1994){Johnston}, {Manchester}, {Lyne},
 {Nicastro}, \& {Spyromilio}}]{1994MNRAS.268..430J}
{Johnston}, S., {Manchester}, R.~N., {Lyne}, A.~G., {Nicastro}, L., \&
 {Spyromilio}, J. 1994, \mnras, 268, 430

\bibitem[{{Khangulyan} {et~al.}(2007){Khangulyan}, {Hnatic}, {Aharonian}, \&
 {Bogovalov}}]{2007MNRAS.380..320K}
{Khangulyan}, D., {Hnatic}, S., {Aharonian}, F., \& {Bogovalov}, S. 2007,
 \mnras, 380, 320

\bibitem[{{Kirk} {et~al.}(2002){Kirk}, {Skj{\ae}raasen}, \&
 {Gallant}}]{2002A&A...388L..29K}
{Kirk}, J.~G., {Skj{\ae}raasen}, O., \& {Gallant}, Y.~A. 2002, \aap, 388, L29

\bibitem[{{Kong} {et~al.}(2011){Kong}, {Huang}, {Tam}, \&
{Hui}}]{2011ATel.3111....1K}
{Kong}, A.~K.~H., {Huang}, R.~H.~H., {Tam}, P.~H.~T., \& {Hui}, C.~Y. 2011, The
Astronomer's Telegram, 3111, 1

\bibitem[{{Lyubarsky} \& {Kirk}(2001)}]{2001ApJ...547..437L}
{Lyubarsky}, Y. \& {Kirk}, J.~G. 2001, \apj, 547, 437

\bibitem[{{Melatos} {et~al.}(1995){Melatos}, {Johnston}, \&
 {Melrose}}]{1995MNRAS.275..381M}
{Melatos}, A., {Johnston}, S., \& {Melrose}, D.~B. 1995, \mnras, 275, 381

\bibitem[{{Michel}(1994)}]{1994ApJ...431..397M}
{Michel}, F.~C. 1994, \apj, 431, 397

\bibitem[{{Moskalenko} \& {Strong}(2000)}]{2000ApJ...528..357M}
{Moskalenko}, I.~V. \& {Strong}, A.~W. 2000, \apj, 528, 357

\bibitem[Negueruela et al.(2011)]{2011arXiv1103.4636N} Negueruela, I., 
Rib{\'o}, M., Herrero, A., Lorenzo, J., Khangulyan, D., 
\& Aharonian, F.~A.\ 2011, arXiv:1103.4636 

\bibitem[{{P{\'e}tri}(2008)}]{2008sf2a.conf..259P}
{P{\'e}tri}, J. 2008, in SF2A-2008, ed. {C.~Charbonnel, F.~Combes, \&
 R.~Samadi}, 259--+

\bibitem[{{P{\'e}tri}(2009)}]{2009A&A...503...13P}
{P{\'e}tri}, J. 2009, \aap, 503, 13

\bibitem[{{P{\'e}tri}(2011)}]{2011MNRAS.412.1870P}
{P{\'e}tri}, J. 2011, \mnras, 412, 1870

\bibitem[{{P{\'e}tri} \& {Kirk}(2005)}]{2005ApJ...627L..37P}
{P{\'e}tri}, J. \& {Kirk}, J.~G. 2005, \apjl, 627, L37

\bibitem[{{P{\'e}tri} \& {Lyubarsky}(2007)}]{2007A&A...473..683P}
{P{\'e}tri}, J. \& {Lyubarsky}, Y. 2007, \aap, 473, 683

\bibitem[{{Sierpowska} \& {Bednarek}(2005)}]{2005MNRAS.356..711S}
{Sierpowska}, A. \& {Bednarek}, W. 2005, \mnras, 356, 711

\bibitem[{{Sierpowska-Bartosik} \& {Torres}(2007)}]{2007ApJ...671L.145S}
{Sierpowska-Bartosik}, A. \& {Torres}, D.~F. 2007, \apjl, 671, L145

\bibitem[{{Sierpowska-Bartosik} \& {Torres}(2008)}]{2008APh....30..239S}
{Sierpowska-Bartosik}, A. \& {Torres}, D.~F. 2008, Astroparticle Physics, 30,
 239

\bibitem[{{Stairs} {et~al.}(2000){Stairs}, {Lyne}, \&
 {Shemar}}]{2000Natur.406..484S}
{Stairs}, I.~H., {Lyne}, A.~G., \& {Shemar}, S.~L. 2000, \nat, 406, 484

\bibitem[{{Tam} {et~al.}(2011){Tam}, {Huang}, {Takata}, {Hui}, {Kong}, \&
 {Cheng}}]{2011arXiv1103.3129T}
{Tam}, P.~H.~T., {Huang}, R.~H.~H., {Takata}, J., {et~al.} 2011, ArXiv e-prints 1103.3129

\bibitem[Tavani \& Arons(1997)]{1997ApJ...477..439T} Tavani, M., \& Arons, J.\ 1997, \apj, 477, 439 

\bibitem[van Soelen 
\& Meintjes(2011)]{2011MNRAS.tmp...17V} van Soelen, B., \& Meintjes, P.~J.\ 2011, \mnras, 17 

\bibitem[{{Wang} {et~al.}(2004){Wang}, {Johnston}, \&
 {Manchester}}]{2004MNRAS.351..599W}
{Wang}, N., {Johnston}, S., \& {Manchester}, R.~N. 2004, \mnras, 351, 599

\end{thebibliography}

\appendix

\onecolumn
\section{Anisotropic Inverse Compton emission from a power-law
distribution of electrons}

In this appendix, we give some details about the analytical
expressions used to compute the spectrum arising from anisotropic
Inverse Compton scattering by a power-law distribution of high energy
leptons with index~$p$ (Eq.~(\ref{eq:FD})).

The scattering rate for an anisotropic and mono-energetic target
photon field with $\vec n$ the direction of propagation of these
incoming photons and $\varepsilon_1$ their energy, normalized to the
electron rest mass energy $m_e c^2$, is given according to
\cite{2000ApJ...528..357M} by
\begin{eqnarray}
\label{eq:SpectreICANISO}
\frac{dN}{dt \, d\varepsilon_2}(\gamma, \varepsilon_1,
\varepsilon_2, \zeta) & = & \frac{3}{8} \, 
\frac{\sigma_T \, c}{\varepsilon_1 \, (\gamma-\varepsilon_2)^2} \,
\left[  2 - 2 \, \dfrac{\varepsilon_2}{\gamma} \, \left(
    \dfrac{1}{\varepsilon_1'} + 2 \right) + \dfrac{\varepsilon_2^2}{\gamma^2} \, \left(
    \dfrac{1}{\varepsilon_1'^2} + \dfrac{2}{\varepsilon_1'} + 3
  \right) - \dfrac{\varepsilon_2^3}{\gamma^3} \right] \\
\varepsilon_1' & = & \gamma \, \varepsilon_1 \, ( 1 - {\bf \beta}
\cdot {\bf n} ) \\
& = & \gamma \, \varepsilon_1 \, ( 1 + \beta \, \cos \zeta ) \\
\varepsilon_2 & \leq & \dfrac{2\,\gamma\,\varepsilon_1'}{1 + 2 \, \varepsilon_1'} < \gamma
\end{eqnarray}
where $\gamma$ is the Lorentz factor of the electron, $\vec\beta$ its
3-velocity, $\zeta$ the angle between photon direction and lepton
velocity ($\zeta=0$ for head-on collision) and $\varepsilon_2$ the
up-scattered photon energy (also normalized to the electron rest mass
energy). All useful quantities are measured in the observer frame
except for $\varepsilon_1'$ which represents the incoming photon
normalized energy as measured in the electron rest frame. We checked
that this formula gives identical results to Eq.~(A.5) of
\citet{2008A&A...477..691D} in the ultra-relativistic limit $\gamma\gg
1$. In this limit, $\frac{dN}{dt \, d\varepsilon_2}$ is a polynomial
in $1/\gamma$. Integration over the particle distribution (i.e. over
$\gamma$) can be performed analytically when the distribution is a
power-law with an integer or half-integer spectral index $p$.

The spectrum is given by
\begin{eqnarray}
\label{eq:Integrale}
\frac{dI}{dt \, d\varepsilon_2}(\varepsilon_1, \varepsilon_2, \zeta)
& = & K_e \, \varepsilon_2 \, \int_{\gamma_1}^{\gamma_2}
\frac{dN}{dt\,d\varepsilon_2} \, \gamma^{-p}\, d\gamma
\end{eqnarray}
with 
\begin{eqnarray}
\label{eq:LimitesPrimitives}
\gamma_1 & = & \textrm{Max} \left( \frac{\varepsilon_2}{2} \, \left[ 1 + \sqrt{ 1 +
      \frac{2}{\varepsilon_1\,\varepsilon_2\,(1+\cos\zeta)} } \right],
  \gamma_{\rm min} \right) \\
\gamma_2 & = & \textrm{Max} \left( \frac{\varepsilon_2}{2} \, \left[ 1 + \sqrt{ 1 +
      \frac{2}{\varepsilon_1\,\varepsilon_2\,(1+\cos\zeta)} } \right],
  \gamma_{\rm max} \right)
\end{eqnarray}
In the case $p=2$ used in our work, we find
\begin{equation}
\label{eq:SPECTREICANISO2}
\int_{\gamma_1}^{\gamma_2} \frac{dN}{dt\,d\varepsilon_2} \, \gamma^{-2}\, d\gamma = 
\frac{1}{1280 \gamma ^5 \pi  {\varepsilon_1}^3 (\gamma
  -{\varepsilon_2}) {\varepsilon_2}^5} \times \nonumber \\
\end{equation}
\begin{eqnarray}
( 3 \left(60 (\gamma -{\varepsilon_2}) (\log (\gamma )-\log (\gamma -{\varepsilon_2}))
  \gamma ^5+{\varepsilon_2} \left(-60 \gamma ^5+30 {\varepsilon_2} \gamma ^4+10
    {\varepsilon_2}^2 \gamma ^3+5 {\varepsilon_2}^3 \gamma ^2+3 {\varepsilon_2}^4 \gamma
    +2 {\varepsilon_2}^5\right)\right) \sec ^4\left(\frac{\zeta }{2}\right) & + & \nonumber \\
10 \gamma 
{\varepsilon_1} {\varepsilon_2} ({\varepsilon_2}-\gamma ) \left(12 (\log (\gamma
  -{\varepsilon_2})-\log (\gamma )) \gamma ^4+{\varepsilon_2} \left(12 \gamma ^3+6
    {\varepsilon_2} \gamma ^2+4 {\varepsilon_2}^2 \gamma +3
    {\varepsilon_2}^3\right)\right) \sec ^2\left(\frac{\zeta }{2}\right) & - & \nonumber \\
\left. 10 \gamma 
{\varepsilon_1}^2 {\varepsilon_2}^2 ({\varepsilon_2}-\gamma ) \left(12 (\log (\gamma
  -{\varepsilon_2})-\log (\gamma )) \gamma ^4+{\varepsilon_2} \left(12 \gamma ^3+6
    {\varepsilon_2} \gamma ^2-4 {\varepsilon_2}^2 \gamma +3
    {\varepsilon_2}^3\right)\right))
 \right|_{\gamma_1}^{\gamma_2} & &
\end{eqnarray}
\label{lastpage}

\end{document}